# Superdiffusion dominates intracellular particle motion in the supercrowded cytoplasm of pathogenic *Acanthamoeba castellanii*

Julia F. Reverey[1], Jae-Hyung Jeon[2], Han Bao[1], Matthias Leippe[3], Ralf Metzler[4,5], Christine Selhuber-Unkel[1*]

[1] *Institute for Materials Science, Biocompatible Nanomaterials, Christian-Albrechts-Universität zu Kiel, Kaiserstr. 2, D-24143 Kiel, Germany.*

[2] *School of Physics, Korea Institute for Advanced Study, Seoul 130-722, Republic of Korea.*

[3] *Zoological Institute, Comparative Immunobiology, Christian-Albrechts-Universität zu Kiel, Olshausenstr.40, D-24098 Kiel, Germany.*

[4] *Institute of Physics & Astronomy, University of Potsdam, D-14776 Potsdam-Golm, Germany.*

[5] *Department of Physics, Tampere University of Technology, FI-30101 Tampere, Finland.*

[*] *Corresponding author; email address: cse@tf.uni-kiel.de*

**Abstract**

Acanthamoebae are free-living protists and human pathogens, whose cellular functions and pathogenicity strongly depend on the transport of intracellular vesicles and granules through the cytosol. Using high-speed live cell imaging in combination with single-particle tracking analysis, we show here that the motion of endogenous intracellular particles in the size range from a few hundred nanometers to several micrometers in *Acanthamoeba castellanii* is strongly superdiffusive and influenced by cell locomotion, cytoskeletal elements, and myosin II. We demonstrate that cell locomotion significantly contributes to intracellular particle motion, but is clearly not the only origin of superdiffusivity. By analyzing the contribution of microtubules, actin, and myosin II motors we show that myosin II is a major driving force of intracellular motion in *A. castellanii*. The cytoplasm of *A. castellanii* is supercrowded with intracellular vesicles and granules, such that significant intracellular motion can only be achieved by actively driven motion, while purely thermally driven diffusion is negligible.

**Introduction**

Intracellular motion is an essential process for a multitude of vital functions, such as cell motility, cell division, and phagocytosis. For the active transport of cargo inside a cell, the interplay of cytoskeletal filaments and molecular motors plays a key role. Examples include myosin motors on actin filaments and kinesin motors on microtubules[1]. An additional role has been attributed to Brownian motion and subdiffusion[2,3]. Both are frequently observed in intracellular motion of single molecules[4,5] and of endogenous and endocytosed particles[6-8], but can also be found for particles moving in *in vitro* biopolymer networks[9-11], which are viscoelastic at intermediate timescales. The type of diffusion present in the intracellular space is strongly determined by the intracellular architecture of the cell, mostly by the arrangement and density of the cellular cytoskeleton and by the action of molecular motors. Theoretically, the existence of subdiffusion has been explained by the presence of obstacles that hinder particles from carrying out normal diffusion[12-16]. With increasing obstacle density, the diffusion exponent decreases significantly[17]. This effect has also been observed in living cells: during the cell cycle, cytoskeletal elements are rearranged, polymerized, and depolymerized. Such structural changes have enormous effects on the diffusive behavior of macromolecules[18] and endogenous granules[19]. Due to its relatively simple, well-understood architecture and its non-motile behavior, fission yeast *S. pombe* is an excellent model system to study different theoretical aspects and concepts of subdiffusion[20] in living cells, e.g. weak ergodicity breaking[7] and the mean maximal excursion method[3].

In contrast to intracellular motion in non-motile *S. pombe* cells, the situation is very different in motile cells. There, intracellular particle motion is always superimposed by the locomotion of the cell body. In spite of such large-scale movements, subdiffusion has been reported to still be present inside motile cells, e.g. for endothelial cells[21], for the social amoeba *Dictyostelium discoideum*[22], and for keratinocytes[23].

Here, we focus on intracellular motion in *Acanthamoeba castellanii,* a free-living amoeba that is often found in soil and water reservoirs. Acanthamoebae are of considerable medical relevance[24,25], as some *Acanthamoeba* species are highly pathogenic. These amoebae are the causative agents of granulomatous amebic encephalitis and amebic keratitis[26], which are difficult to cure. Pathogenic amoebae destroy host tissues and kill host cells in a contact-dependent reaction, in which the release of cytolytic factors, such as pore-forming toxins or metalloproteinases, is involved[27-30]. For the highly pathogenic *Acanthamoeba culbertsoni*, a unique pore-forming protein, termed acanthaporin, has recently been comprehensively characterized[31]. Prior to host cell destruction, acanthamoebae form close contact with host cells. Subsequently, endogenous granules move to the contact site where they presumably release cytolytic factors into the extracellular space. Intracellular motion is therefore a crucial prerequisite for acanthamoeba pathogenicity. Understanding the basic mechanisms involved in intracellular motion and transport inside acanthamoebae is hence essential to get a complete picture of the pathogenicity of these protozoan parasites.

This investigation is particularly interesting, as the intracellular space in acanthamoebae appears densely packed with different types of endogenous particles, such as digestive and contractile vacuoles, smaller vesicles, and granules (Fig. 1). The largest vacuole in acanthamoebae is the contractile vacuole, which is responsible for osmotic regulation. Digestive vacuoles contain precipitates and amorphous substrates[32]. Lysosomes store various hydrolytic enzymes and are responsible for the digestion of endocytosed food and other foreign particles inside the phagolysosome, e.g. bacteria, fungi, and viruses[33].

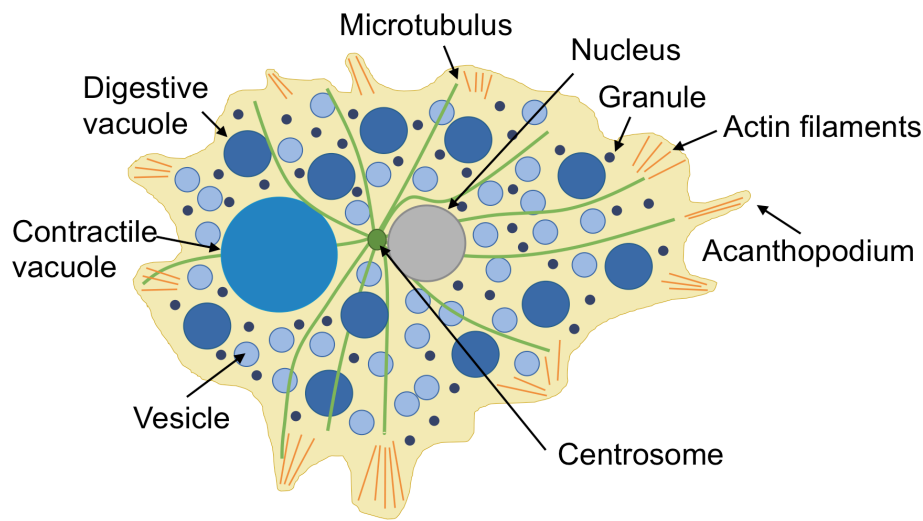

*Fig. 1: Schematic picture of the intracellular architecture of an acanthamoeba, showing that many different types of vacuoles and granules densely fill the cytoplasm.*

Actin, microtubules, and myosin motors are also ubiquitous in acanthamoebae. Microtubules originate from the so-called centrosome near the nucleus and spread out in a dense 3D network throughout the cell body[34]. Baumann and Murphy showed that mitochondria co-localize with the microtubules in *A. castellanii* and that the movement of mitochondria and small particles (< 1µm) is inhibited by colchicine, a microtubule depolymerizing substance[35]. Actin plays a key role in the contraction of the contractile vacuole, as myosin IC is located around the vacuole prior to contraction[36-39]. Myosin I is mainly present near the rear edge of the acanthamoebae and in the filopodia, which are here typically called acanthopodia[39-42], whereas myosin II is present within the whole cytoplasm and is concentrated in the cell cortex[37]. The state of macromolecular crowding in biological cells with proteins and cytoskeletal structures was previously referred to as `superdense[6]. Here we use the notion of `supercrowded´ volume to point out that in the *A. castellani* cells considered here we additionally find a large amount of vacuoles of several microns in size.

Here, we focus on investigating the intracellular motion of endogenous particles, such as vesicles and granules, in *A. castellanii* under different experimental conditions. We emphasize the relation between intracellular motion, acanthamoeba locomotion and the contribution of cytoskeletal elements. Intracellular motion and acanthamoeba locomotion are investigated by high-speed live cell imaging and theoretical analysis concepts, including mean-square displacement (MSD) and velocity autocorrelation function analyses. Our results show a predominance of superdiffusion that cannot be explained by the *a priori* statement that intracellular particles are swept along with the locomoting acanthamoeba. We observe a striking involvement of myosin II motors, which turns out to be essential for maintaining motion in the supercrowded intracellular volume of *A. castellanii*.

**Results**

Following the tracks of individual particles in the intracellular space reveals valuable information about the physical properties of their intracellular motion. Fig. 2A shows a representative phase contrast image of an *A. castellanii* trophozoite. The white halo surrounding the cell body arises from the ellipsoidal shape of the acanthamoeba, which only slightly flattens during attachment. 2D tracks of individual intracellular particles from the acanthamoeba in Fig. 2A are plotted in Fig. 2B. The size of the tracked particles ranged from a few hundred nanometers to several micrometers. For almost all tracked particles we observe a consistent directionality in motion. This directionality appears to be defined by the direction of movement of the acanthamoeba, so that intracellular particles are swept along with the cell body. Still, we find a few exceptions from this strongly directional motion, showing that not only cell drift causes intracellular particles to move, but that additional mechanisms for individual particle motion must exist, e.g. by the contribution of molecular motors. Analyzing the TA MSDs (see Methods section) of the particle tracks revealed that the diffusion exponent $\alpha$ is close to 2 for almost all particle tracks (Fig. 3A, Table 1). Some single particles show normal diffusive behavior, subdiffusive motion is negligible.

In order to investigate the influence of acanthamoeba locomotion on the trajectories of individual particles, we assume that particle motion in a moving acanthamoeba is the sum of the drift due to acanthamoeba locomotion and its additionally induced intracellular motion $\vec{r}_I(t) = (x_I(t), y_I(t))$, such that $\vec{r}(t) = \vec{r}_c(t) + \vec{r}_I(t)$. Here, acanthamoeba locomotion is characterized by the position of its centroid $\vec{r}_c(t)$, which is determined from the outline of the acanthamoeba after image segmentation (Suppl. Fig. S7). $\vec{r}(t)$ are the raw particle trajectories and $\vec{r}_I(t)$ represents the particle motion relative to the centroid of the acanthamoeba. The trajectories $\vec{r}_I(t)$ have a very different appearance (Fig. 2C) with a much less pronounced directionality than the raw trajectories $\vec{r}(t)$ shown in Fig. 2B. Still, our TA MSD analysis reveals that intracellular particle motion relative to the centroid of the acanthamoeba is superdiffusive (Fig. 3B). This effect can also be demonstrated by plotting $<r^2(\Delta)>_t/\Delta$, $<r^2(\Delta)>_t/\Delta^2$ (Suppl. Fig. S8) and stacked TA MSDs (Suppl. Fig. S9).

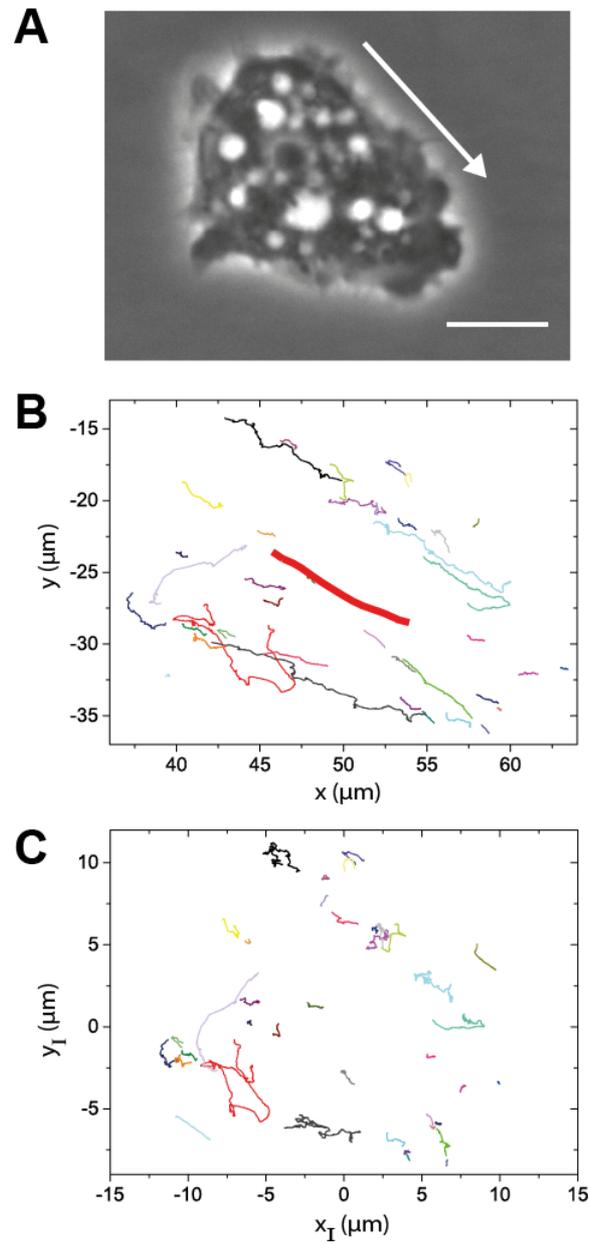

*Fig. 2: Intracellular particle motion in an acanthamoeba and its relation to locomotion. (A) Phase contrast image of an* A. castellanii *trophozoite crawling on a surface. Scalebar: 10 µm. The arrow indicates the direction of acanthamoeba locomotion. (B) Particle trajectories tracked inside the acanthamoeba shown in A. Most tracks follow the direction of movement of the acanthamoeba centroid (bold red line). However, not all particles obey this trend. (C) Particle trajectories $x_I(t)$, $y_I(t)$ of intracellular particles relative to the position of the acanthamoeba centroid. No significant directionality of particle motion can be observed.*

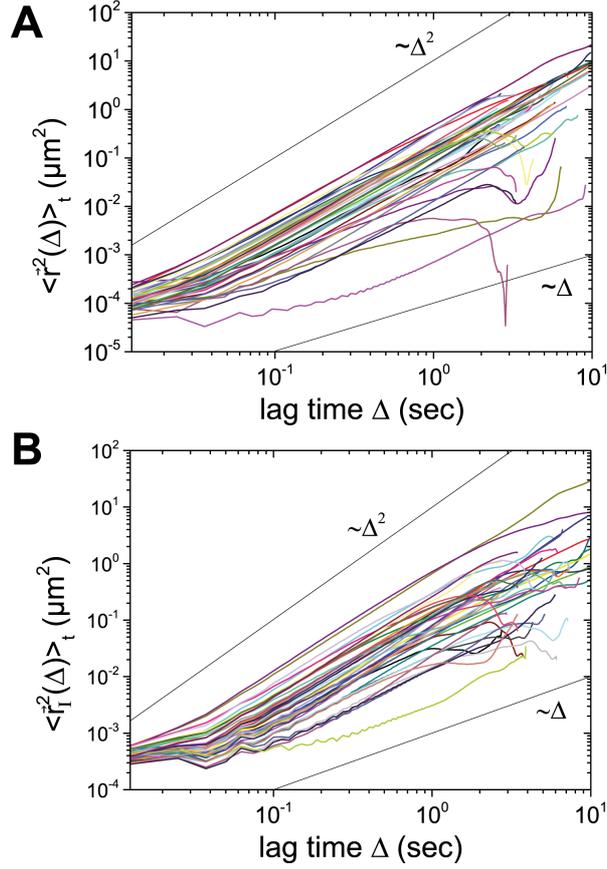

*Fig. 3: Representative TA MSD plots of intracellular particle trajectories in* A. castellani. *(A) TA MSD of intracellular particles. Most of the particles show superdiffusive motion with a diffusion exponent close to α ~ 2. (B) TA MSD of particle trajectories relative to the centroid of the acanthamoeba. Colors shown here are consistent with the colors used for the tracks in Fig. 3A.*

From individual particle trajectories $\vec{r}(t)$ we evaluated their corresponding velocity autocorrelation functions, as described in the Methods section and in the Supplementary Information. Fig. 4A depicts the theoretical curves of the drift-FBM model defined below for three different cases: (i) if intracellular motion is subdiffusive, the velocity autocorrelation has a minimum after the time interval $\delta t$ and then relaxes towards its saturation value; (ii) if intracellular motion is normally diffusive, essentially there is no relaxation after $\delta t$; (iii) if particle motion is superdiffusive, the velocity autocorrelation is always positive at all times and its relaxation occurs above the saturation value. Note that the velocity autocorrelation eventually saturates to a constant $V_d^2/(V_d^2 + 4K_\alpha \delta t^{\alpha-2})$ so that the saturation value is always positive irrespective of the drift direction, and for $\delta t \neq 1$ it has a different value if $\alpha$ is varied albeit the drift velocity is the same. Such a constant drift velocity is in good agreement with the experimentally determined motion of the acanthamoeba centroid (Suppl. Fig. S7C). In the FBM model, the long-time relaxation towards its saturation decays as $\sim |t|^{\alpha-2}$. This means that the relaxation of superdiffusive motion is always slower than that of subdiffusion[43,44].

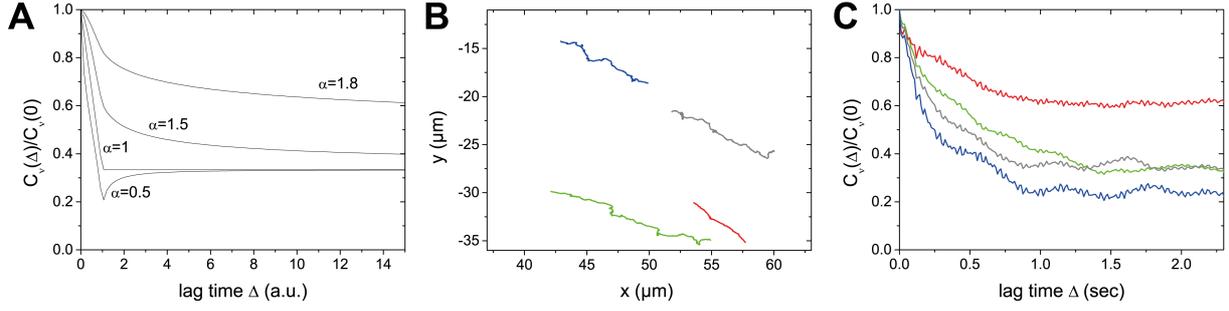

*Fig. 4: Velocity autocorrelation analysis of particle trajectories. (A) Theoretical velocity autocorrelation functions (VACF) for FBM-like intracellular motion assuming a constant drift of the amoeba for α=0.5, 1, 1.5, and 1.8. Parameters: $\delta t = 1, K_\alpha = 1, V_d = 1$. (B) 2D plot (x(t),y(t)) of arbitrarily chosen, long intracellular particle trajectories. (C) VACFs of the particle trajectories shown in B. These profiles are representative VACFs of intact acanthamoebae. Color codes in B and C are the same.*

Fig. 4B shows raw particle trajectories used in the VACF analysis. Fig. 4C depicts the corresponding VACF curves. Here, we observe the following: (i) The obtained pattern of VACF curves is similar to the case of superdiffusive FBM considered in Fig 4A. The curves continuously decay from unity with time, without a dip, towards a positive constant. Consequently, the intracellular motion of particles in the acanthamoeba studied is superdiffusive, presumably due to the activity of molecular motors and strong intracellular fluid flow (see supplementary movie). This directional persistency in intracellular diffusion has a characteristic time of about a second. (ii) The fact that the VACF saturates towards a constant value tells us that the acanthamoeba indeed moves with an almost constant drift as in the assumption of our model. Such a constant movement appears to be in the time window of a few seconds. (iii) There is a trajectory-to-trajectory fluctuation in the saturation value of the VACF. Although a constant drift motion was observed, such a fluctuation is expected because the saturation value of VACF depends on the diffusion exponent α and the diffusivity of the particles, as well as on the drift speed.

In order to investigate the role of myosin II for intracellular particle motion, we treated acanthamoebae with different concentrations of blebbistatin, a myosin II specific inhibitor [45]. At concentrations smaller than 50 μM, blebbistatin did not inhibit intracellular motion completely: at concentrations above 50 μM, most acanthamoebae detached from the surface (data not shown). Accordingly, we chose 50 μM for intracellular particle tracking experiments. Fig. 5A shows a phase contrast image of a blebbistatin-treated *A. castellanii*. Intracellular structures, such as vacuoles, granules, and intracellular organelles are not as clearly visible as in an intact acanthamoeba. Right after the addition of blebbistatin, acanthamoeba locomotion disappeared almost completely, and also intracellular dynamics were drastically reduced (Fig. 5B). Even for these comparably long tracks of more than 20 sec duration, intracellular particle motion is almost negligible. Specifically, particles exhibit subdiffusion where $\langle \vec{r}^2(\Delta) \rangle_t \sim \Delta^\alpha$ with $\alpha \sim 0.2$ at $\Delta \lesssim 1$ sec, and almost normal diffusion at $\Delta \gtrsim 1$ sec (Fig. 5C). The normally diffusive motion for these confined particles reflects the fact that the intracellular components in acanthamoebae move altogether in a diffusive way at long times. Active intracellular motion was not observed for any of the tracks, revealing a dramatic change of intracellular dynamics due to myosin II inhibition, from superdiffusion in intact acanthamoebae to almost completely suppressed intracellular motion in myosin II inhibited acanthamoebae.

Notably, some time after addition of blebbistatin we observed a change in cell morphology (Fig. 5D) and intracellular movement restarted (Fig. 5E). However, particle trajectories look at first glance different from trajectories of intact *A. castellanii* (Fig. 2B and Fig. 2C), as they are not yet very directional. The TA MSDs show that intracellular motion of many particles is superdiffusive in this situation (Fig. 5F).

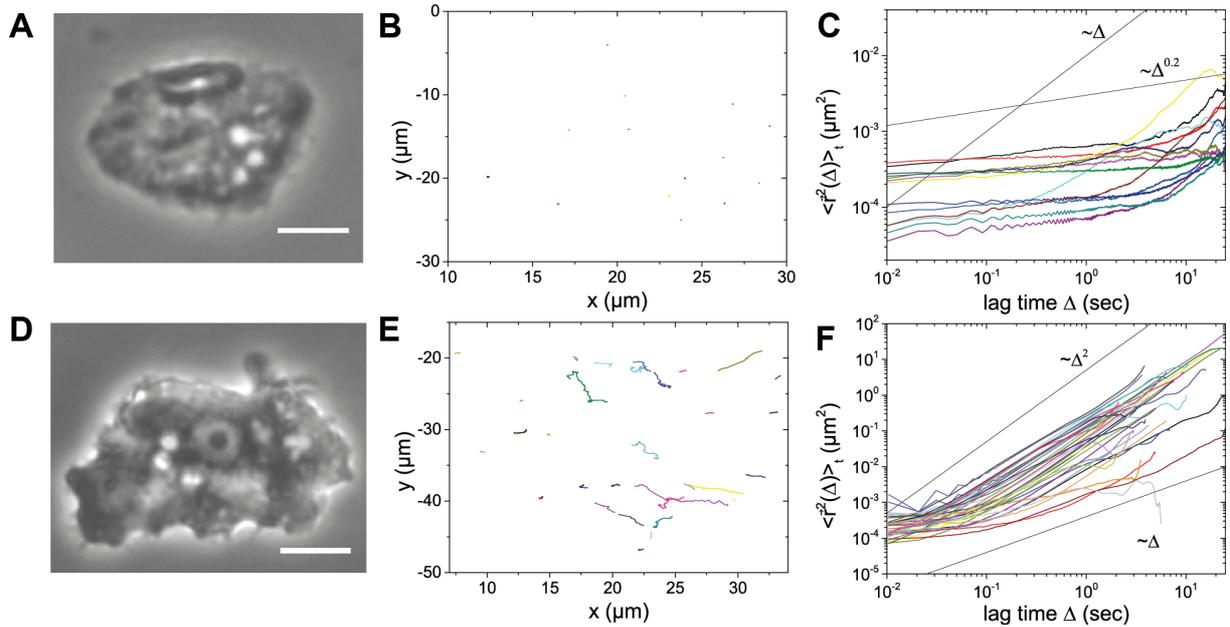

*Fig. 5: Analysis of intracellular particle motion in blebbistatin-treated acanthamoebae. (A) Phase contrast image of* A. castellanii *treated with blebbistatin for inhibiting myosin II motors. Scalebar: 10 μm  (B) Particle trajectories indicate that intracellular motion is almost eliminated. (C) Strongly confined motion is observed in the TA MSDs. (D) Intracellular particle tracks for the same amoeba 1 h after the addition of blebbistatin. Intracellular motion (E) and superdiffusion (F) can be observed.*

In Fig. 6A we plot representative VACF curves from single particle trajectories of a blebbistatin-treated acanthamoeba. It shows that there is almost no drift in the motion (as the relaxation decays to almost zero) and there is a dip (negative correlation) after Δ=0.1 sec. Such a VACF curve is typical for subdiffusive particle motion without drift. The experimental VACF curve is very similar to the theoretical curve of FBM with α=0.2 (black solid line). As demonstrated in theoretical and experimental studies[10,13,46-48] viscoelastic fluids induce strong anti-correlation of particle displacements, leading to FBM-like anomalous diffusion[15,49]. Thus, the negative value in the VACF curve shown here suggests that the highly restricted passive motion of particles is connected to the viscoelastic nature of the supercrowded intracellular fluid.

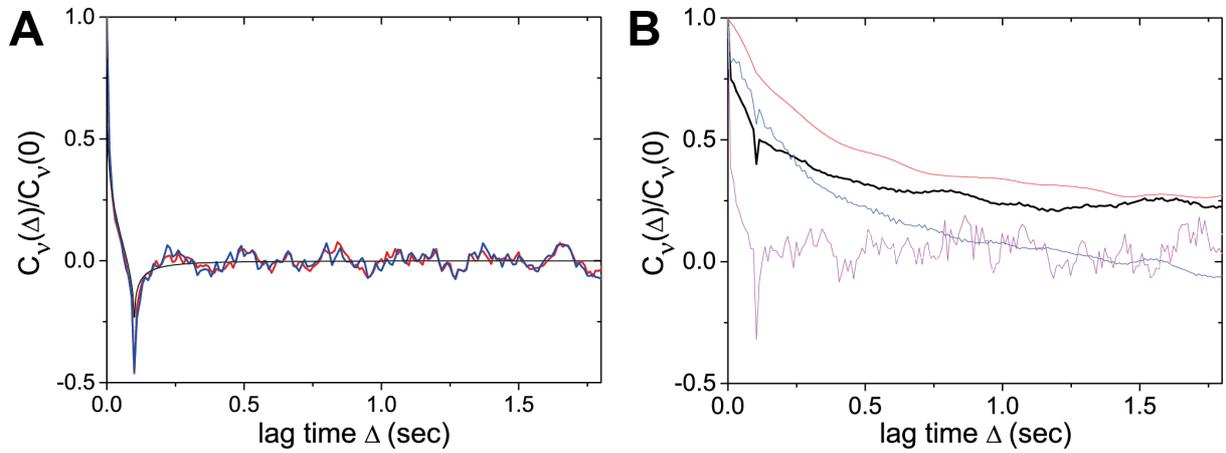

*Fig. 6: Velocity autocorrelation function for intracellular particles in a blebbistatin-treated acanthamoeba (A) and after 1 h of recovery from inhibitor treatment (B). (A) Blue and red lines correspond to VACFs of particle trajectories. The black line denotes the theoretical VACF curve for FBM with α=0.2. (B) Red, blue and grey lines show different types of VACFs received from the trajectory analysis. The black line denotes the average of 37 analyzed trajectories.*

A clear difference is observed when investigating the velocity autocorrelation function for particle trajectories in acanthamoeba after 1 h of recovery. In Fig. 6B we plot three such representative VACFs that clearly show that the particles carry out various types of diffusive motion. This result is consistent with the trajectories (Fig. 5E) and the TA MSDs (Fig. 5F). The average VACF shows superdiffusive motion plus drift, in agreement with our observation that the acanthamoeba is motile again.

Due to the strong effect of myosin II on intracellular motion, actin was our next target of investigation. We used latrunculin A to inhibit the polymerization of the actin filaments and a concentration of 5 μM latrunculin A turned out to be optimal, as higher concentrations often induced an immediate detachment of acanthamoebae. At lower concentrations acanthamoebae still moved on the surface and no pronounced effect of latrunculin A was observed. At concentrations higher than 5 μM, most acanthamoebae detached from the surfaces. A few minutes after latrunculin A addition we typically observe that cell locomotion was slowed down. The shape of acanthamoebae became relatively smooth with much fewer protrusions compared to intact cells (Fig. 7A). Within acanthamoebae, particles still moved and many small particles (size < 500 nm) could be observed close to the membrane, where normally large particles, presumably vacuoles, are crowded together and make the optical identification of tiny particles impossible. However, many acanthamoebae detached during the experiments, presumably because actin is a key player for their adhesion. As shown in Fig. 7B, latrunculin A does not stop intracellular motion and intracellular flow, even though locomotion of the acanthamoeba itself is stopped, and there is no distinct directionality in the particle trajectories. In particular, the TA MSD analysis proves that the particles mainly carry out superdiffusive motion with a diffusion exponent of about $\alpha \sim 1.6$ (Fig. 7C). Velocity autocorrelations for selected trajectories (Fig. 7D) also show positive correlations, in agreement with the superdiffusive TA MSD and the observation of intracellular flow.

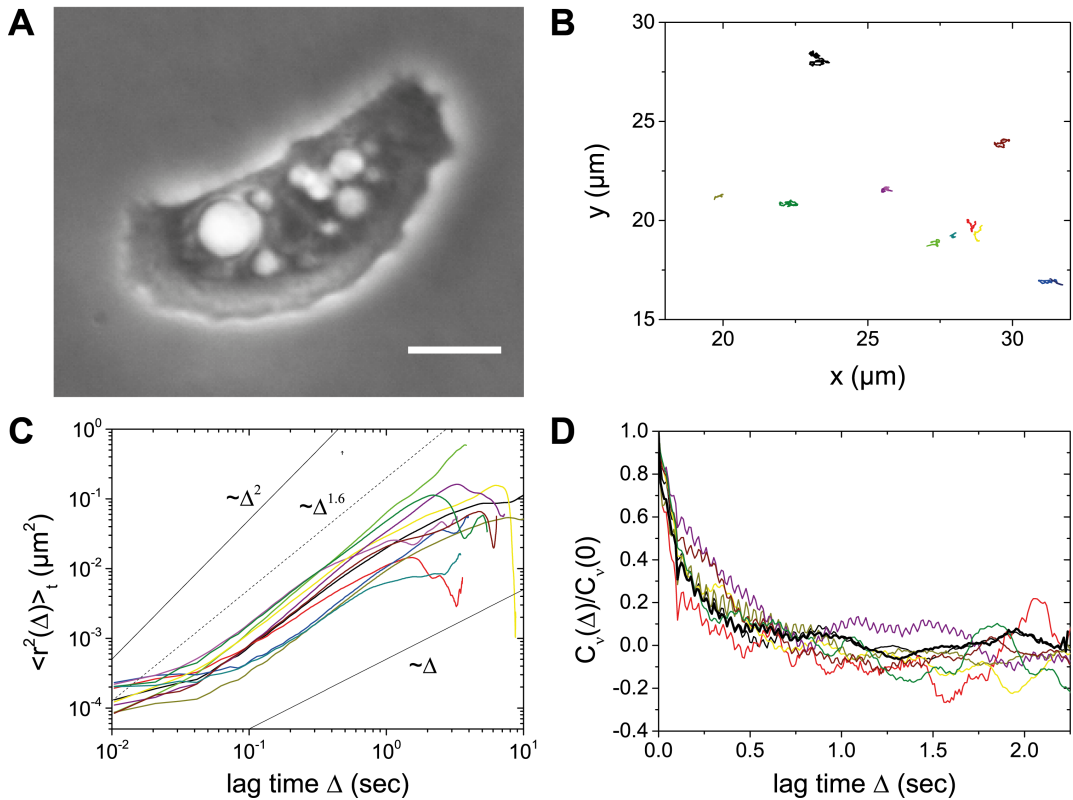

*Fig. 7: Intracellular particle motion in an acanthamoeba treated with 5 μM latrunculin A. (A) Phase contrast image. Scalebar: 10 μm. (B) 2D intracellular particle trajectories of the acanthamoeba shown in A. The tracks do not show strong directionality. (C) TA MSDs reveal superdiffusion. (D) Velocity autocorrelation function for selected representative trajectories. The bold black line denotes the average of 13 analyzed trajectories.*

Microtubules are suggested to be key players for intracellular particle motion, also in amoebae[50]. Acanthamoebae treated with 40 µM nocodazole for microtubule depolymerization have a comparably rough shape (Fig. 8A) and their locomotion is drastically reduced compared to intact acanthamoebae. At lower concentrations, such a pronounced effect of nocodazole was not observed. Within acanthamoebae there were many vacuoles, but only a few tiny particles were visible. In contrast to latrunculin A treatment, the adhesion of nocodazole-treated acanthamoebae to the surface was not impaired. The motion of acanthamoebae was drastically affected by nocodazole, but not completely inhibited. Intracellular motion was observed (Fig. 8B), but no clear directionality was found for particle motion. Comparable to latrunculin A, the TA MSDs here reveal diffusion exponents of about $\alpha \sim 1.7$ (Fig. 8C).

The difference between latrunculin A and nocodazole treatments becomes particularly visible in the velocity autocorrelation functions. Under the influence of nocodazole, the velocity autocorrelation shows a constant drift motion. This can be a result of the remaining slow drift motion of the acanthamoeba during the experiment (Fig. 8D).

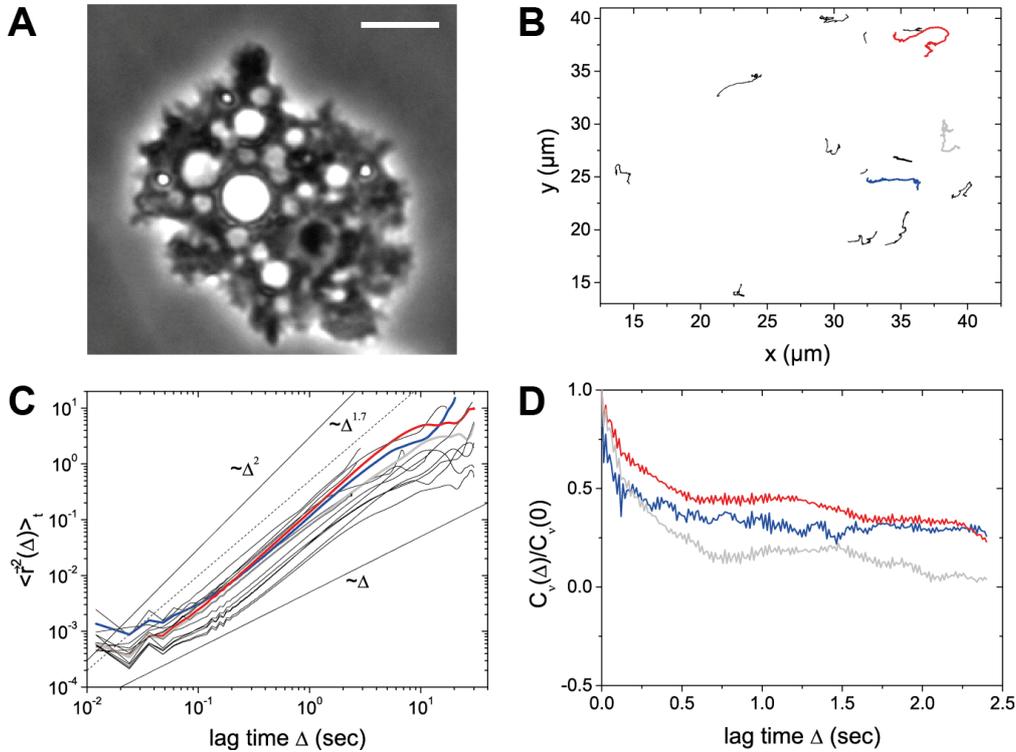

*Fig. 8: Influence of nocodazole on intracellular motion in acanthamoebae. (A) Particle tracks in an acanthamoeba treated with 40 µM nocodazole. Scalebar: 10 µm  (B) 2D intracellular particle trajectories of the acanthamoeba shown in A. In contrast to the intact acanthamoeba, the tracks do not show strong directionality. (C) TA MSD analysis reveals superdiffusion. (D) VACF of the red, blue, and grey tracks shown in B and C.*

In order to clearly visualize the effect of different drugs, we summarized the distributions obtained for the diffusion exponent α in the different drug treatment situations (Fig. 9). Whereas a striking effect is obtained between blebbistatin and all other situations, the difference between latrunculin A and nocodazole treatment is negligible and the values of α are for both treatments only slightly smaller than in the case of intact acanthamoebae. In table 1 averaged values of α are shown and information on the size of the datasets is given.

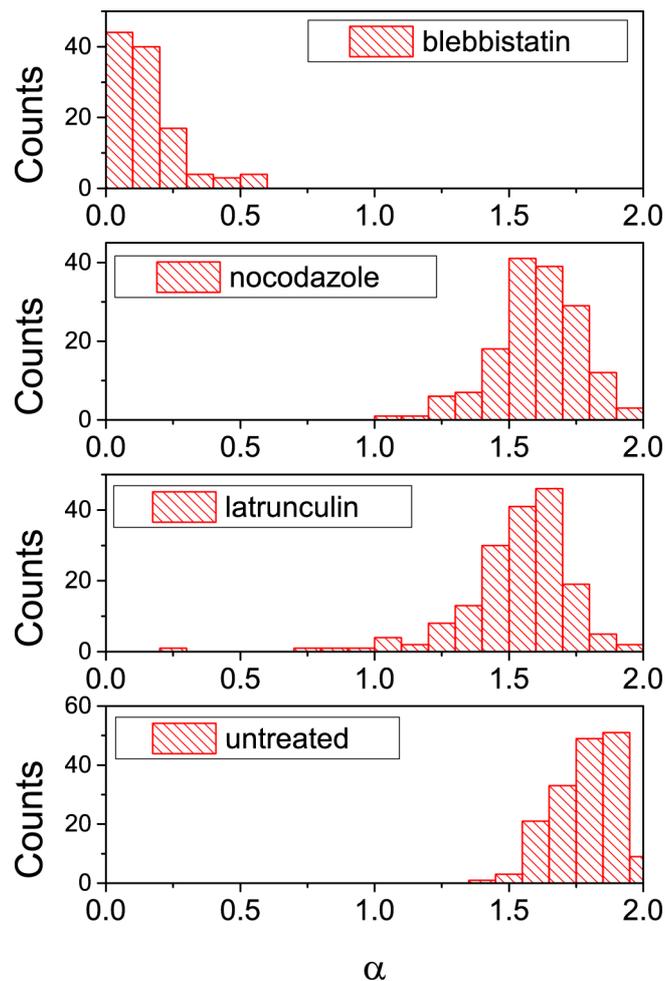

*Fig. 9: Influence of different treatments on the diffusion exponent α. The influence of the different treatments is clearly visible, particularly the extremely strong impact of blebbistatin, giving rise to the pronounced subdiffusion with 0<α<0.5. In all other cases shown here we observe superdiffusion with α >1.*

**Discussion**

We have used time-lapse imaging and particle tracking algorithms to investigate the role of locomotion, cytoskeletal elements, and myosin II motors on intracellular particle movement in the protozoon and human pathogen *A. castellani* by analyzing mean-square displacements and velocity autocorrelations. The experiments were complex because acanthamoebae move extremely fast and exhibit quick shape changes[25].

For intact acanthamoebae we found that superdiffusion dominates the motion of endogenous intracellular particles. This is very different from previous results on non-motile and shape-conserved *S. pombe* yeast cells, where subdiffusion dominates intracellular particle motion[3,7,8,19]. Surprisingly, in *A. castellanii* even the motion of intracellular particles relative to the acanthamoeba centroid is superdiffusive (Fig. 3B). At first glance this result contradicts recently published work by Kinneret et al., who showed that the motion of intracellular quantum dots relative to the centroid of a keratinocyte is subdiffusive[23]. However, it has to be considered that acanthamoebae are, in contrast to keratinocytes, densely packed with vacuoles and granules. As already noted, acanthamoebae also frequently change their shape during locomotion, whereas keratinocytes hardly do. Accordingly, one may assume that intracellular motion in acanthamoebae relies on different principles to ensure the transport of intracellular cargo, in spite of their dense packing. Hence, we believe that active superdiffusive motion is a strategy of acanthamoebae to allow particle movement in its supercrowded intracellular volume. This becomes particularly evident in the experiments in which myosin II motors were blocked, as intracellular motion and acanthamoeba locomotion vanished consistently in the TA MSD and in the VACF analysis, leading to a diffusion exponent $\alpha \sim 0.2$. These results show that the intracellular volume is considerably more supercrowded than the bacterial cytoplasm[6].

A well-known result of inhibition of actin polymerization is a decreased diffusion exponent $\alpha$ both for non-motile[8] and motile cells[22,51]. We also observed a slight decrease of $\alpha$ in our experiments, with a decrease from in average about 1.8 to about 1.5 and 1.6 in response to latrunculin A and nocodazole treatments, respectively. This also means that intracellular motion stays superdiffusive in spite of inhibiting the polymerization of cytoskeletal elements. Hence, cytoskeletal elements themselves play a minor role for intracellular motion in acanthamoebae, in agreement with the results on the cytoplasm of *E. coli* bacteria[6]. Instead, myosin II is an essential driving force to maintain intracellular particle motion. In contrast to intracellular particle tracking in breast cancer cells, where only a minor influence of latrunculin A and blebbistatin treatments on the diffusion exponent was detected[52], we here clearly observe an impact of both inhibiting myosin II and actin polymerization. Interestingly, our diffusion exponent in the case of untreated acanthamoebae is larger than the one detected in previous publications on superdiffusive intracellular particle motion[52,53]. Therefore it is even more striking that motion appears to be almost completely blocked by myosin II inhibition in acanthamoeba. As we studied the motion of particles endogenous to acanthamoeba, our results are intrinsically different from other results on superdiffusive intracellular motion, where the motion of engulfed artificial particles[52,53], also under the influence of mechanical disturbances[54], was investigated.

Our data show that the TA MSD analysis and the VACF analysis are consistent and support our assumption that intracellular particle motion in *A. castellanii* is dominated by superdiffusion. In particular, the VACFs reveal that intracellular motion follows an FBM-type Gaussian process and is therefore consistent with the drift-FBM model (see Methods). We note that superdiffusion also occurs in vacancy-dominated, confined crowded systems[55-57].

In conclusion, intracellular particle motion in *A. castellanii* is a prime example for the importance of superdiffusion in a supercrowded, motile, and spatially dynamic volume. We have shown that superdiffusivity not only arises from the fast migration of acanthamoebae, but also from superdiffusive motion relative to the centroid of an acanthamoeba. Myosin II turned out to play a crucial role for the functionality of intracellular and cellular transport systems: if myosin II was inhibited, acanthamoebae neither migrated nor exhibited intracellular particle motion. Although inhibiting actin polymerization and depolymerizing microtubules changed the overall cell shape and could almost stop acanthamoeba migration, intracellular particle motion was not completely eliminated. Inhibiting myosin II caused strongly confined intracellular motion, so that we consider myosin II motors to be key players for maintaining such motion. Brownian motion and subdiffusion appear to play negligible roles for intracellular transport in acanthamoebae because the extreme crowding would limit all transport-related biological functions.

With regard to amoebic pathogenicity, the dominant role of active particle motion agrees well with the notion that it relies – at least in part – on the fast and targeted intracellular transport of toxic factors stored in lysosome-like cytolytic granules through a supercrowded intracellular volume. Our study may serve as a starting point for further investigations on the interplay between crowding, diffusion, and active transport in cells in general and their role for the pathogenicity of amoebic parasites.

## Methods

### Culture of acanthamoebae

*A. castellanii* (ATTC 30234) were cultured in PYG medium at room temperature in 75 ml tissue culture bottles (Sarstedt, Germany) as described previously[58,59]. Prior to experiments, about $10^5$ freshly harvested acanthamoebae (*A. castellanii*) per ml medium were seeded on a flat surface (tissue culture bottle or six-well plate, Sarstedt AG & Co., Nümbrecht, Germany) and incubated at room temperature for 30 min to 1 h to ensure their attachment to the tissue culture surface.

### Time-lapse imaging

*Acanthamoeba* locomotion and intracellular motion were observed with an inverted phase contrast microscope (Olympus CKX41 and IX81, Olympus Deutschland GmbH, Hamburg, Germany) with either a LCACHN 40X PHP, a LUCPLANFLN 40X PH2 or a UPLFLN 60X (long distance) objective (all from Olympus). Sequences having a length of at least 30 sec were recorded with a high-speed camera (Hamamatsu C9300, Hamamatsu, Japan) at frame rates between 80 and 99 frames per second (fps). Exposure times chosen for image recording are not expected to cause errors in the analysis of particle trajectories[60].

### Inhibition of actin polymerization

Latrunculin A inhibits the polymerization of actin filaments as it binds to the subunits of actin filaments and avoids further reaction of the subunits to longer filaments[61]. For our experiments, Latrunculin A (Sigma-Aldrich Chemie GmbH, Munich, Germany) was dissolved in DMSO (Carl Roth GmbH & Co. KG, Karlsruhe, Germany) at a 1 mM stock solution concentration. Acanthamoebae were cultured as described above and latrunculin A was added to reach final concentrations of 2 μM, 5 μM, 7.5 μM or 10 μM. High-speed sequences were recorded a few minutes after latrunculin A addition and again after 1-2 h to ensure acanthamoeba survival.

### Inhibition of microtubule polymerization

Nocodazole was used to inhibit microtubule polymerization. It binds to the subunits of the microtubules and thus prevents polymerization[62]. Stock solutions of nocodazole (Sigma-Aldrich Chemie GmbH, Munich, Germany) in DMSO (Carl Roth GmbH & Co. KG, Karlsruhe, Germany) were prepared at concentrations of 1 mM. Acanthamoebae were cultured as described above and nocodazole was added to reach final concentrations of 10 µM, 20 µM and 40 µM. Live-cell imaging was carried out as described before and was started a few minutes after addition of nocodazole. Acanthamoebae viability was verified 1-2 h after the experiment.

**Inhibition of myosin II motors**

For inhibiting myosin II motors, (-)-blebbistatin was used. The reagent binds to ATPase, which is a catalyzer for ATP hydrolysis and hence blocks the myosin heads in their unbound state as it slows down ATP hydrolysis, especially the release of phosphate[45]. (-)-blebbistatin (Sigma-Aldrich, Germany) was dissolved in DMSO (Roth, Germany) at 1 mM concentration. (-)-blebbistatin was added to the medium of the cultured acanthamoebae to reach different concentrations (50 µM, 60 µM, and 70 µM). Right after the addition of (-)-blebbistatin, time-lapse imaging was started. Here, we employed for the experiments a phase contrast microscope (Olympus IX81, Olympus Deutschland GmbH, Hamburg, Germany) with a 40× (LCACHN 40X PHP) objective. After a period of 15 min to 1 h the effect of (-)-blebbistatin decreased and time-lapse sequences were recorded again to prove the viability of the cells by monitoring intracellular motion and locomotion. Acanthamoebae had completely normal appearance after 1-2 h of relaxation from the drug treatment.

**Evaluation of time-lapse image sequences**

For tracking the position of intracellular particles, an automated image processing algorithm (Polyparticletracker) specially designed for tracking endogenous cellular particles was used in Matlab (The MathWorks, Natick, MA)[63,64]. Image segmentation and acanthamoeba centroid detection were carried out with a Matlab program of our own based on built-in algorithms for segmenting cells in phase contrast images. Centroid position was fitted by a polynomial function (see Supplementary Information) in order to remove strong fluctuations arising from short-term acanthopodia extensions and retractions.

**Mean-square displacement analysis**

The time-averaged mean-square displacement (TA MSD) of the tracked particles was calculated from

$$\langle \vec{r}^2(\Delta) \rangle_t = \frac{1}{T-\Delta} \int_0^{T-\Delta} dt [\vec{r}(t+\Delta) - \vec{r}(t)]^2 \quad (1)$$

with $\vec{r}(t) = (x(t), y(t))$. $\Delta$ denotes the lag time in seconds and $T$ is the overall measurement time. For typical timescales, the TA MSD is characterized by $\langle \vec{r}^2(\Delta) \rangle_t \sim 4K_\alpha \Delta^\alpha$ where α is the diffusion exponent with $0 < \alpha \leq 2$ and $K_\alpha$ is the generalized diffusion coefficient of dimension [cm$^2$/sec$^\alpha$]. For $\alpha < 1$ particles carry out subdiffusion, for $\alpha = 1$ normal diffusion (Brownian motion), and for $\alpha > 1$ superdiffusion. In order to determine the distribution of α values, we fitted the linear region of the TA MSDs in the double-logarithmic plot between Δ=0.1 s and Δ=1 s using MATLAB.

**Analysis of the velocity autocorrelation function**

The time-averaged velocity autocorrelation function (VACF) was calculated using

$$C_v(\Delta) = \frac{1}{T-\Delta-\delta t} \int_0^{T-\Delta-\delta t} \vec{V}_{\delta t}(t+\Delta) \cdot \vec{V}_{\delta t}(t) dt \quad (2)$$

for a given trajectory $\vec{r}(t)$ and the time interval $\delta t$. $\vec{V}_{\delta t}(t)$ refers to the average velocity at time *t* over the time interval $\delta t$, which is defined as $\vec{V}_{\delta t}(t) = (\vec{r}(t+\delta t) - \vec{r}(t))/\delta t$. All results for the velocity autocorrelation function presented below are this time-averaged quantity. In our analysis, the time interval $\delta t$ was fixed to $\delta t = 10\Delta t_{frame} \approx 0.12\ s$. By studying the relaxation profile of the velocity autocorrelation we can obtain information on the drift of an acanthamoeba as well as on the stochastic properties of intracellular motion. In particular, we investigated the case where intracellular motion follows a Fractional Brownian motion (FBM)-type Gaussian process that can be obtained for subdiffusive, normal diffusive, or superdiffusive motion[43,44,65]. As demonstrated in theoretical and experimental studies[10,13,46,47], FBM-like anomalous diffusion typically occurs in viscoelastic environments, which give rise to long-time correlation in spatial displacements. It has been shown that intracellular diffusion is in many cases governed by FBM, at least on the relevant intermediate time scales, due to the viscoelastic nature of cytoplasmic solutions[5,7,20,66-68]. Based on this knowledge, we consider here the case that intracellular particles follow FBM and the acanthamoeba hosting the particles moves with a constant drift speed $V_d$. More details are given in the Supplementary Information.

**Acknowledgements**

We thank Heidrun Ließegang for her advice in culturing acanthamoebae, Ali Raza for support with image segmentation, Brook Shurtleff and Steven Huth for proofreading the manuscript. R. M. acknowledges support by the Academy of Finland (Suomen Akatemia) within the Finland Distinguished Professorship program. M.L. was supported by the "Cluster of Excellence Inflammation at Interfaces", Cluster laboratory X, of the German Research Foundation (DFG). C.S. and J.R. acknowledge support from the German National Academy of Sciences Leopoldina by grant LPD 9901/8-164, as well as from the DFG through the Emmy Noether program (grant SE-1801/2-1), and the SFB 677 (project B11).

**Author contributions**

J. F. R. and H. B. conducted experiments and analyzed data. J.-H. J. and R. M. analyzed data and designed theory, M. L. provided acanthamoebae and contributed to the design of the study, C. S. designed the experiments and analyzed data. All authors contributed to writing the manuscript.

**Additional information**

Supplementary information and supplementary videos accompany this paper.

There are no competing financial interests.

**Table 1:** Average diffusive exponents for different drug treatments and numbers of analyzed acanthamoebae and intracellular particles.

| inhibitor | $\alpha$ | # (amoebae) | # (particles) |
|---|---|---|---|
| none | 1.79 ± 0.12 | 8 | 167 |
| latrunculin A | 1.53 ± 0.21 | 8 | 174 |
| nocodazole | 1.60 ± 0.16 | 5 | 157 |
| blebbistatin | 0.15 ± 0.13 | 11 | 112 |